\documentclass[prl, reprint]{revtex4-1}
\usepackage[usenames,dvipsnames,svgnames,table]{xcolor}
\usepackage{hyperref}
\usepackage{amsmath, amsfonts,amssymb, mathtools} 
\hypersetup{colorlinks=true,  linkcolor=NavyBlue, citecolor=PineGreen,urlcolor=cyan}
\usepackage{physics}
\usepackage{graphicx}

\usepackage[mathscr]{euscript}

\renewcommand{\H}{\mathscr{H}} 

\graphicspath{{./Figures/}}

\renewcommand{\a}{\alpha}

\renewcommand{\r}[1]{\textrm{#1}}
\newcommand{\avg}[2][]{\left< #2 \right>_{#1}} 

\begin{document}

\title{Universal behavior of the full particle statistics of one-dimensional Coulomb gases with an arbitrary external potential}

\author{Rafael D\'{i}az Hern\'{a}ndez Rojas} 
\affiliation{Dipartimento di Fisica,  Universit\`{a} di Roma ``La Sapienza'', P.le A. Moro 5, I-00185 Rome, Italy}
\author{Christopher Sebastian Hidalgo Calva} 
\affiliation{ Institute of Physics, UNAM, P.O. Box 20-364, 01000 Mexico City, Mexico}
\author{Isaac P\'{e}rez Castillo} 
\affiliation{Department of Quantum Physics and Photonics, Institute of Physics, UNAM, P.O. Box 20-364, 01000 Mexico City, Mexico}
\affiliation{London Mathematical Laboratory, 14 Buckingham Street, London WC2N 6D, United Kingdom}

\begin{abstract}
 We present a complete theory for  the full particle statistics of the positions of  bulk and extremal particles in a one-dimensional Coulomb Gas (CG) with an arbitrary potential, in the typical and large deviations regimes.  Typical fluctuations are described by a universal function which depends solely on general properties of the external potential. The rate function controlling large deviations is, rather unexpectedly, \emph{not} strictly convex and has a discontinuous third derivative around its minimum for both extremal \emph{and} bulk particles. This implies, in turn, that the rate function cannot predict the anomalous scaling of the typical fluctuations with the system size.  Moreover, its asymptotic behavior for extremal particles differs from the predictions of the Tracy-Widom distribution. Thus many of the paradigmatic properties of the full particle statistics of two-dimensional systems do \emph{not} carry out into their one-dimensional counterparts, hence proving that 1d CG belongs to a different universality class.  Our analytical expressions are thoroughly compared with Monte Carlo simulations showing excellent agreement. 
\end{abstract}

\maketitle

Being universality  one of the pillars of modern theoretical physics, an important   goal is  to understand under which conditions universal properties do emerge in strongly correlated systems, together with their range of validity. Pursuing an answer to this poignant question, classical ensembles of random matrices have become a mathematical laboratory that allows  to test these ideas, as the joint probability density  of its eigenvalues offers  a correlated system   which, moreover, is  simple enough to be amenable to a thorough and rigorous mathematical treatment. This can quite generally be written as
\begin{equation}
\label{eq:pdf eigenvalues}
P(\vb{x}) = C_N e^{- \frac{\beta}{2} \left(V(\vb{x}) - \sum_{i\neq j}\log(\abs{x_i-x_j}) \right) }\qc 
\end{equation}
 with $\vb{x}=\{x_i\}_{i=1}^N$,  $C_N$ is a normalization constant, and $V(\vb{x})$ is a function depending on the particular ensemble  \cite{mehta-rmt,log-gases-book-forrester}.  Physically,   Eq.~\eqref{eq:pdf eigenvalues} can be identified as the two-dimensional Coulomb interacting system  $N$ charged particles, constrained to move along a single direction, with an external potential $V(\vb{x})$,  the so-called Dyson's log gas \cite{dyson-log-gas}. Using path integral methods, this Coulomb fluid picture has been used to study  the asymptotic  behaviour of the statistics of extreme and bulk eigenvalues in several classical ensembles \cite{Dean2006,Vivo2007,Dean2008,Vergassola2009,Majumdar2009,Katzav2010,Majumdar2011,Ramli2012,Majumdar2012b,Allez2013,Majumdar2013,Majumdar2014,Perez2014,Perez2014b,Camacho2015,Perez2016}.  In particular, the statistics of  extremal  eigenvalues  follows a universal behavior  governed by the Tracy-Widom (TW) distribution \cite{Tracy1994,Tracy1996}, while the typical fluctuations of bulk eigenvalues scale logarithmically with the system size rather than linearly \cite{Gustavsson2005,Orourke2010}. The main physical reason behind these, and other findings, has been well established by now and corresponds to abrupt changes, or phase transitions, on  the different mechanisms governing the  statistical fluctuations. 
 
The validity of this ubiquitous statistical behaviour has been further explored in other correlated systems inspired mainly in ensembles of random matrices  by either considering non-invariant ensembles \cite{Metz2016,Perez2018,Perez2018b} or by probing correlated systems similar to that  in Eq. \eqref{eq:pdf eigenvalues} but with a different inter-particle interaction. An important result on the latter  was considered in \cite{Cunden2017}, where it was shown that there is a discontinuity in the third derivative of the rate function describing the large deviations of the extremal particle in a CG confined by an arbitrary external central potential for any dimension, thus pinpointing an universal third order phase transition, according to the Ehrenfest criterion. Moreover, in Ref.~\cite{Dhar2017} it has been shown that when the 1d CG is subjected to an external harmonic potential,  the  statistics of the rightmost particle exhibits a different distribution from the TW around its typical value. These recent results make clear that the study of CG with different dimensionality may provide either deeper understanding of their shared universal properties or give rise to new behaviors that contrast the traditional, celebrated ones of RMT.

To show that many universal features of the CG are indeed sensible to the physical dimensions of the system, we present here the complete solution of the full particle statistics of the 1d case with an arbitrary external potential, obtaining exact expressions for  both  the typical and  large fluctuations regimes.  To be specific, we  consider a Hamiltonian of the form 
\begin{equation}
\label{eq:H}
\H(\vb{x}) = N^2 \sum_{i=1}^N v(x_i) - N \a \sum_{i<j}^{1,N} \abs{x_i-x_j}\, ,
\end{equation}
where the choice of the powers of $N$ ensures that we have a non-trivial contribution in the thermodynamic limit.  Clearly,   to have a confined configuration $v(x)$ must be a convex function,  but an upper bound on $\alpha$  might also be required to guarantee that $v(x)$ dominates over the electrostatic repulsion and  an equilibrium particle density $\rho_\r{eq}(x)$ is attained \cite{Chafai2014,Dhar2017,Cunden2017}. The choice of the interparticle potential  corresponds to the  so called ``jellium'' model or one-dimensional one-component plasma and has been studied on distinct scenarios as many relevant quantities can be calculated exactly \cite{Lenard1961,Baxter1963,Choquard1981,Dean2010,Trizac2015}.   As the Hamiltonian in Eq.~\eqref{eq:H} is invariant under the permutation of particles, we will henceforth assume that  $x_\r{min}\equiv x_1 \leq x_2 \leq \cdots \leq x_N\equiv x_\r{max}$. Then the optimal position of the $i$-th particle, $x_i^*$ is thus given by \footnote{See Supplemental Information for further details.}:
\begin{equation}
v'(x_i^*) = \frac{\a}{N} (2i - N -1)\,,\quad i=1,\dots,N \,,
\label{eq:x optimal}
\end{equation}
thus in the thermodynamic limit  $\rho_\r{eq}(x)$ has a natural domain $x\in [x_-, x_+]$,  with $v'(x_\pm)=\pm \a$. Depending on the external potential, a restriction on the values of $\alpha$ may be necessary to obtain a physical solution \cite{Note1}. In addition, when $v(x)$ is of class $C^3$, we have that for the Hamiltonian of Eq.~\eqref{eq:H}, the typical fluctuations regime correspond to deviations of order $\mathcal{O}(N^{-1})$, while large deviations  are of order $\mathcal{O}(N^0)$ \cite{Chafai2014,Cunden2017,Dhar2017,Note1}.

To obtain the cumulative distribution function (CDF) of the typical fluctuations of both, the extremal particles  and bulk particles ($x_K,\ 1<K<N$) around their average positions, $\avg{x_i}=x_i^*$, $i=1,\dots,N$, we  write the probability $ \r{Prob}[x_1<x_2<\cdots<x_i<w]= Z_c(w;N)/Z_c(\infty;N)$  ,  for a fixed but arbitrary particle indexed according to $i=cN$ and with  \cite{Dhar2017,Note1}: 
\begin{multline}
Z_{c}(w; N) = N! \int_{-\infty}^w \dd{x_i}\ \prod_{j=1}^{i-1} \int_{-\infty}^{x_{j+1}} \dd{x_j} \\
\times  \prod_{j=i+1}^N \int_{x_{j-1}}^{\infty} \dd{x_j} e^{-\H(\vb{x})}
\end{multline}
By using the Taylor expansion of $\H(\vb{x})$ up to second order, and defining $W_i\equiv N u_i(w-x_i^*)$, $\epsilon_j \equiv N u_j(x_j-x_j^*)$, $\Delta^{(\pm)}_j = \pm Nu_j(x_{j\pm1}^*-x_j^*)$, $y^{(\pm)}_j= \frac{u_{j\mp 1}}{u_{j}}\epsilon_{j} \pm \Delta^{(\pm)}_{j\mp 1}$ and $u_i=\sqrt{v''(x_i^*)}$ the last integral can be approximated as
\begin{multline}
Z_c(w; N) \approx \frac{N!\ e^{-\H(\vb{x}^*)} }{N^N \prod_{j=1}^N u_j}
\int_{-\infty}^{W_i} \dd{\epsilon_i} e^{-\frac12 \epsilon_i^2}\\
\times \qty( \prod_{j=1}^{i-1} \int_{-\infty}^{y_{j+1}^{(+)}}\dd{\epsilon_j} e^{-\frac{1}{2}\epsilon_j^2} ) 
\qty( \prod_{j=i+1}^{N} \int_{y_{j-1}^{(-)}}^{\infty}  \dd{\epsilon_j} e^{-\frac12 \epsilon_j^2}  )  \, .
 \label{eq:Z c general}
\end{multline}

Note that the first (resp. second) multiple integral inside the parenthesis in Eq.~\eqref{eq:Z c general} is proportional to the CDF of the extremal particle, being smaller (resp. greater) than $x_i$, but for a smaller system of size $i-1$ (resp. $N-i$). This suggests that the fluctuations of the bulk particles can be described in terms of the CDF of the extremal ones, \textit{i.e} $c=1$ and $c=0$ \cite{Note1}. To shorten notation let us write  $F_c(W_i(w);N)\equiv Z_c(w;N)/Z_c(\infty;N)$. Then  the statistics of $x_\r{max}$, whose  CDF is $F_1(W;N)$ obeys the following forward differential equation in the thermodynamic limit \cite{Dhar2017,Baxter1963,Note1}:
\begin{equation} 
\label{eq:ode F1}
\dv{F_1(W)}{W} = A_1\ e^{-W^2/2}\ F_1\qty(W + \frac{2\a}{u_+}) \, ,
\end{equation}
where $u_\pm = \sqrt{v''(x_\pm^*)}$ and $A_1$ is a constaint which is fixed upon imposing  boundary conditions  $F_1(W)\to1$ as $W\to\infty$, and $F_1(W)\to0$ as $W\to-\infty$.  An entirely analogous analysis can be made for the statistics of the leftmost particle, corresponding to $F_0$, for which the resulting PDF is determined by a delayed differential equation.\\
The typical fluctuations for  bulk particles are obtained by choosing $K=cN$ and considering the thermodynamic limit while $c$ remains finite. The corresponding CDF, denoted  $F_c(W)$, obeys the following forward and delayed differential equation \cite{Note1}:
\begin{equation} 
\dv{F_c(W)}{W} = A_c\ e^{-W^2/2}\ F_1\qty(W+\frac{2\a}{u_c}) F_0\qty(W-\frac{2\a}{u_c}),
\label{eq:ode F general}
\end{equation}
where  $u_c=\sqrt{v''(x)}$ with $x$ such that $c=\int_{x_{-}}^x \rho_{\rm{eq}}(y) dy$, and $A_c$  can be obtained by requiring $F_c(W)$ to be normalized. 

\begin{figure}[h!]
\includegraphics[width=0.47\linewidth,height=5.4cm]{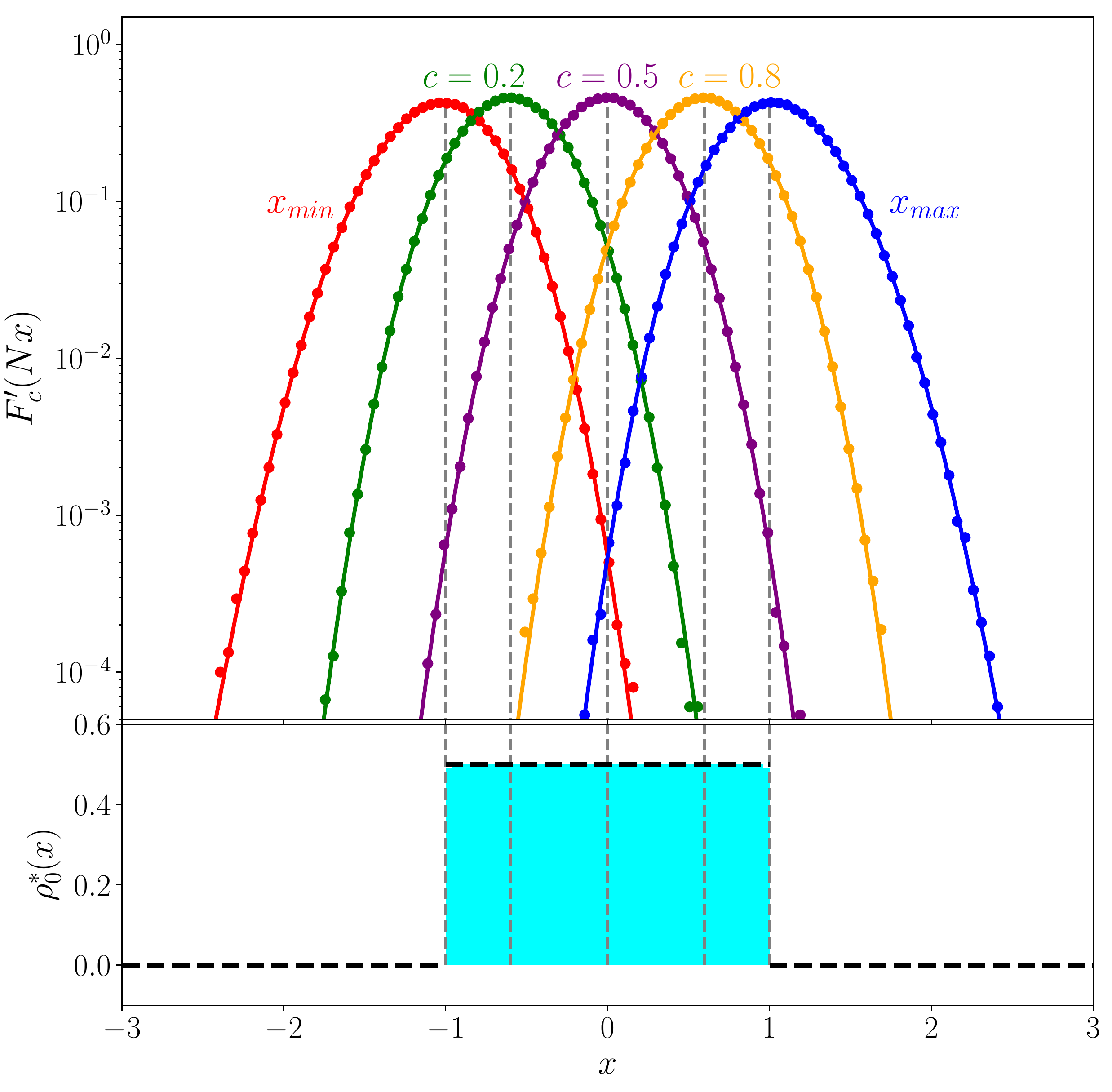}
\includegraphics[width=0.47\linewidth,height=5.4cm]{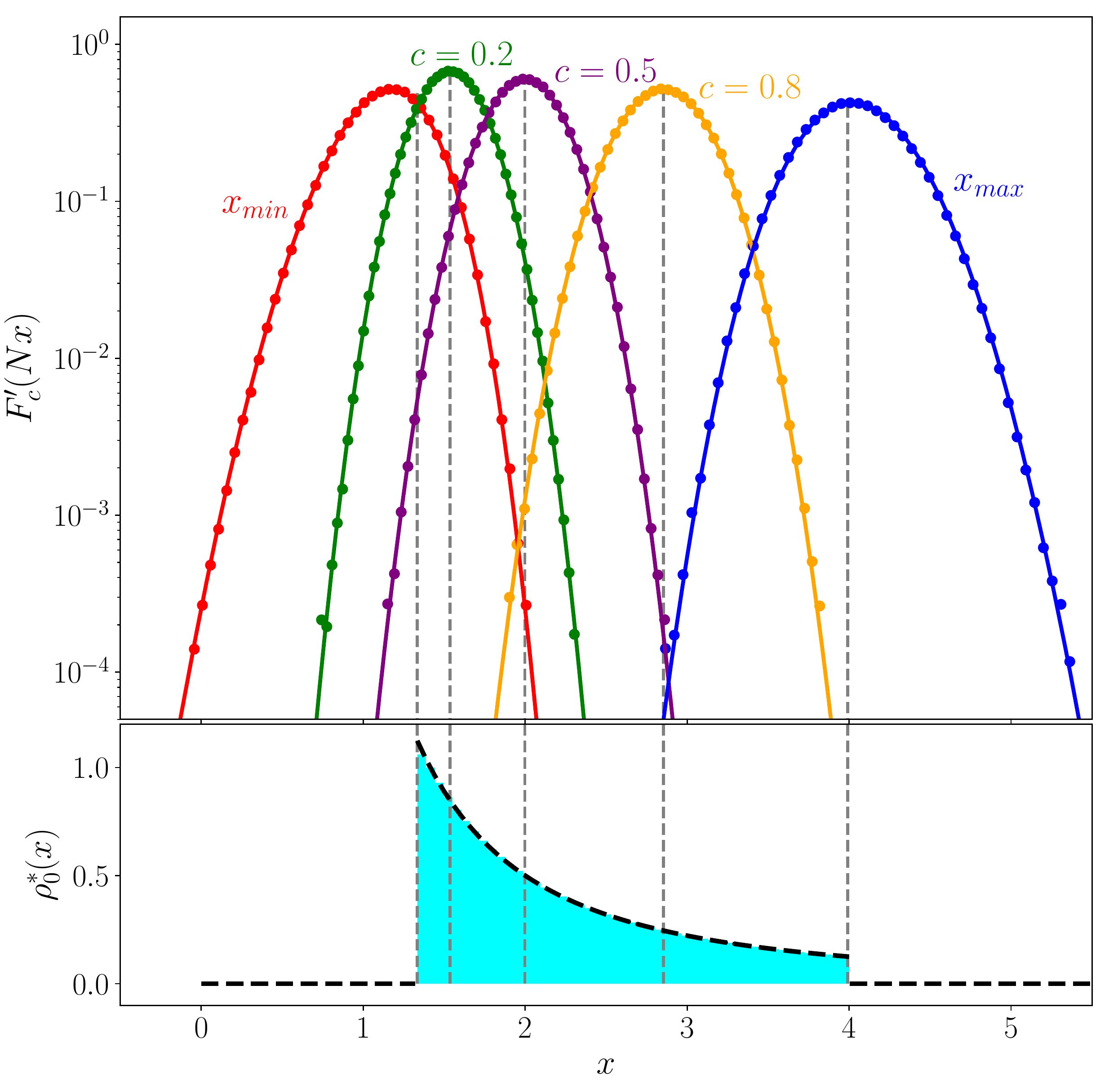}
\caption{Top panels: Comparison of the PDFs $\dv{F_c(W)}{w}$ obtained by solving numerically Eqs.~\eqref{eq:ode F1} and \eqref{eq:ode F general} for different values of $c$ (solid curves) and MC simulations (markers) considering two potentials. Left panel: $v(x)=\frac{x^2}{2}$ that  corresponds to the classical Gaussian ensemble. Right panel:  $v(x)=\frac12(x-a\log x)$, with $a>1$, inspired by the Wishart-Laguerre ensemble.  A  factor $N$ in the argument of $F_c$ has been included to magnify the size of the fluctuations. Bottom panels: Density of the 1d CGs using the same potentials, comparing the MC samples (cyan histogram) with the analytic expression for $\rho_\r{eq}(x)$ of Eq.~\eqref{eq:unconstrained density} (black dashed curve). The dashed vertical lines indicate the position for which the typical fluctuations were calculated. We used $10^6$ MC steps for a system of $N=500$ particles.}
\label{fig:fluctuations xK}
\end{figure}

Eqs.~\eqref{eq:ode F1} and \eqref{eq:ode F general} are the first of our main results, for they provide a complete description of the full set of particles, now indexed according to $c$. Secondly, they show that the joint contribution of the confining potential $v(x)$ as well as the electrostatic interaction is captured succinctly by the constants $\frac{2\a}{u_\pm}$ and $\frac{2\a}{u_c}$. This means that $F_c(W)$ is indeed an \emph{universal function} for  one-dimensional CGs  describing their typical statistics for \textit{any} external convex potential. It is fairly straightforward to show \cite{Note1} that the asymptotic behavior of $F'_1(W)$ is given by
\begin{equation}\label{eq:asymptotic F1}
F'_1(W) \asymp \begin{dcases}
\exp(-W^2/2), & W\to \infty \, ,\\
\exp(-\frac{u_+}{12\a} \abs{W}^3), & W\to -\infty \,.
\end{dcases}
\end{equation}
Notice that the right tail, $W\to \infty$, is rather different from the 2d case governed by the asymptotic TW pdf that decays as $e^{-\frac32 W^{3/2}}$ \cite{Majumdar2014}. Similarly, the asymptotic behavior of the PDF for bulk particles turns out to be
\begin{equation} \label{eq:asymptotic Fc}
F'_c(W) \asymp \exp(-\frac{u_c}{12\a} \abs{W}^3)\qc W\to \pm \infty \, .
\end{equation}
As typically large fluctuations are expected to match atypical small ones, Eq. \eqref{eq:asymptotic Fc} indicates that the rate function is not strictly convex and therefore will be unsuitable to describe the Gaussian-like behavior found in the 2d case \cite{Gustavsson2005,Majumdar2009,Orourke2010,Perez2014,Perez2016}. Moreover, we will see, that the rate function has a 3rd order discontinuity. These results show that the 1d CG belongs to a different universality class than the one determined by the TW distribution. To conclude our analysis of the typical fluctuations regime, we present in Fig.~\ref{fig:fluctuations xK} the results obtained by solving numerically the differential equations above and a comparison with Monte Carlo (MC) simulations using the Hamiltonian of Eq.~\eqref{eq:H} for two paradigmatic potentials of RMT.

To study the large deviation regime for which $\abs{w-x_i}\sim \order{1}$, for any $i=1,\dots,N$ we use the  Coulomb fluid method \cite{Dean2006,Vivo2007,Dean2008,Vergassola2009,Majumdar2009,Katzav2010,Majumdar2011,Ramli2012,Majumdar2012b,Allez2013,Majumdar2013,Majumdar2014,Perez2014,Perez2014b,Camacho2015,Perez2016} to compute $\varrho(c,w)\equiv \r{Prob}[\mathcal{C}_w=c]$, with $ \mathcal{C}_w= \frac1N \sum_{j=1}^N \Theta(w-x_j)$, which corresponds to the probability that \emph{exactly} $cN$ particles have positions smaller than $w$.  We start by writing
\begin{equation}
\varrho(c,w) = \frac{1}{\Omega_0} \int \dd{\vb{x}} p(\vb{x})\ \delta\qty(c- \frac1N \sum_{i=1}^N \Theta(w-x_i) )\, ,
 \label{eq:condicional density}
\end{equation}
with $p(\vb{x})=\frac{1}{\Omega_0}e^{-\H(\vb{x})}$. This can be written as  the following path integral (and two integrals over variables $\mu$ and $\nu$) $\varrho(c,w)= \frac{1}{\Omega_0} \int D[\rho,\mu,\nu] e^{-N^3 \mathcal{S}[\rho, \mu, \nu]}$ with  $\mathcal{S}$ being the action \cite{Note1}

\begin{multline} 
\mathcal{S}[\rho, \mu, \nu] = \int \dd{x} \rho(x) v(x) - \frac{\a}{2} \int \dd{x}\dd{x'} \abs{x-x'} \rho(x)\rho(x')\\
 - \mu \qty(1 - \int\dd{x}\rho(x) ) - \nu \qty(c- \int \dd{x} \Theta(w-x) \rho(x) ) \, . 
 \label{eq:action}
\end{multline}
Here, $\mu$ and $\nu$ are Lagrange Multipliers to enforce normalization in the density $\rho$ and that a fraction $c$ of particles are to the left of $w$, respectively. Similarly, the normalization constant can be written as $\Omega_0 = \int D[\rho_0,\mu_0] e^{-N^3 \mathcal{S}_0[\rho_0, \mu_0]}$ and corresponds, in turn, to a CG without a wall. In the thermodynamic limit both expressions can be evaluated by the saddle-point method obtaining  $\varrho(c,w)\asymp e^{-N^3 \psi(c,w) }$ where
\begin{equation}
\psi(c,w) = \mathcal{S}[\rho^*, \mu^*, \nu^*] - \mathcal{S}_0[\rho^*_0, \mu^*_0]  
\label{eq:psi actions difference}
\end{equation}
is  the rate function. Here $\rho^*(x)$  corresponds physically to the  equilibrium particle density of a system constrained to  have a fraction of particles $c $ to the left of $w$,  while $\rho_0^*(x)$ is the unconstrained equilibrium particle density.  As noted in \cite{Perez2014,Perez2016}, the rate function  $\psi(c,w) $  has a dual role, for it describes the large deviations of  $\mathcal{C}_w$, when $w$ is taken as a parameter or, conversely, the statistics of the $i$-th particle when $\psi$ is viewed as a function of $w$.\\
Noteworthy, the stationarity conditions of $\mathcal{S}$ yield an integral equation that can be solved \emph{exactly} for \emph{any} external potential $v(x)$. The solution for the unconstrained system is \cite{Note1}:
\begin{equation} 
\rho_0^*(x) = \frac{v''(x)}{2\a} \mathbb{I}[x_- \leq x \leq x_+] \,,
\label{eq:unconstrained density}
\end{equation}
where $\mathbb{I}[A]$ is an indicator function, whose value is $1$ whenever condition $A$ is true. The bottom panels of Fig.~\ref{fig:fluctuations xK} show a comparison of this formula with unconstrained MC samplings. 

From a physical perspective, we expect the constrained density to be similar to $\rho_0^*(x)$ since by placing a barrier at $w$ there will only be a \emph{local} density rise in its vicinity. This is a consequence of the regularity of the 1d inter-particle potential when particles overlap, in contrast with its 2d counterpart. Hence we have that once the wall is present, there will be an accumulation of particles next it, whose magnitude depends on the fraction of the particles ``pushed'' by it, compared to the one of the unconstrained system. The latter one, denoted as $c^*(w)$, is easily calculated by integrating $\rho_0^*(x)$ up to $w$. When $w\in[x_-,x_+]$ we have $c^*(w) = \frac{v'(w)+\a}{2\a}$ and the constrained equilibrium density results into \cite{Note1}:

\begin{multline}
\rho^*(x) = \frac{v''(x)}{2\a} \left(\mathbb{I}[x_- \leq  x \leq a] + \mathbb{I}[b <  x \leq x_+] \right)\\
 + \abs{c-c^*(w)}\ \delta(w-x) \, .
 \label{eq:density}
\end{multline}
Here, the several parameters involved in Eq. \eqref{eq:density} are defined as follows: $x_0$ is such that $v'(x_0)=\a(2c-1)$; when $c>c^*(w)$  we must take  $a=w$ and $b=x_0$, while for $c<c^*(w)$ we have instead that  $a=x_0$ and $b=w$ . For $c=c^*(w)$, the wall becomes ineffective, and therefore $x_0=w$,  recovering the unconstrained solution $\rho^*_0(x)$. Finally,  for $w<x_-$  (resp. $w>x_+$) we have that $c^*=0$ (resp. $c^*=1$) and similar expressions for $\rho^*(x)$ apply.  The fact that the resulting equilibrium density has, in general, an infinitely sharp peak at $w$ as well as a non-compact and bounded support resembles some well known results for the spectral densities of RMT \cite{Dean2006,Vivo2007,Dean2008,Vergassola2009,Majumdar2009,Katzav2010,Majumdar2011,Ramli2012,Majumdar2012b,Allez2013,Majumdar2013,Majumdar2014,Perez2014,Perez2014b,Camacho2015,Perez2016}. However, in those systems the effect of introducing the wall significantly modifies the unconstrained density and obtaining an analytical expression for $\rho^*(x)$ is only possible in few, exemplary cases. Surprisingly, this is not longer true for 1d CG, where we have found the equilibrium density for any convex potential and any fraction of particles to the left of $w$. It is important to mention that so long as $v(x)$ is strong enough to dominate as $\abs{x}\to \infty$, the equilibrium densities of Eqs.~\eqref{eq:unconstrained density} and \eqref{eq:density} are the unique minimizers of the corresponding actions \cite{Chafai2014}, while the convexity of the potential assures that they are non-negative functions. 

\begin{figure}[h!]
\centering
\includegraphics[width=0.494\linewidth, height=4.5cm]{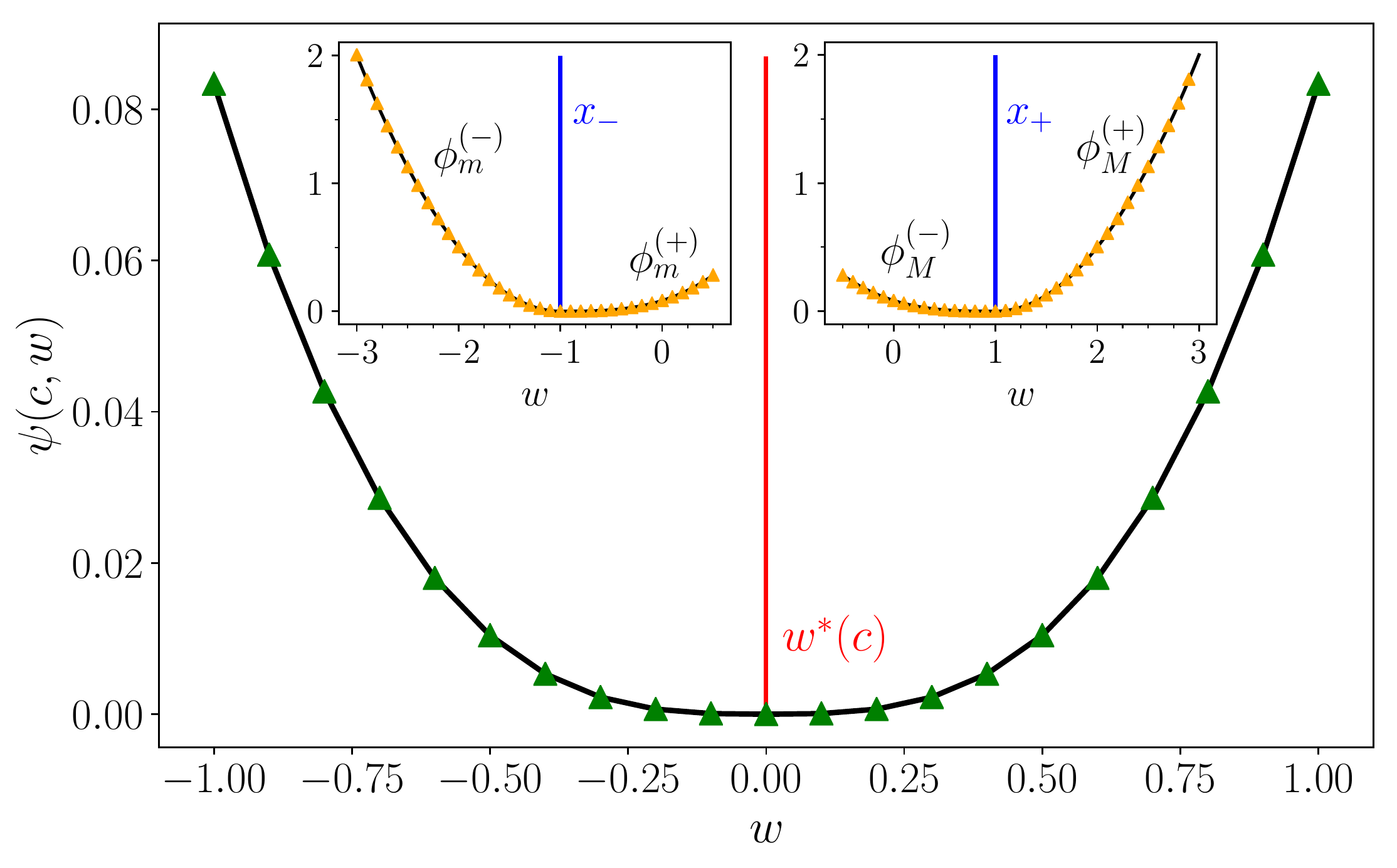} 
\includegraphics[width=0.494\linewidth, height=4.5cm]{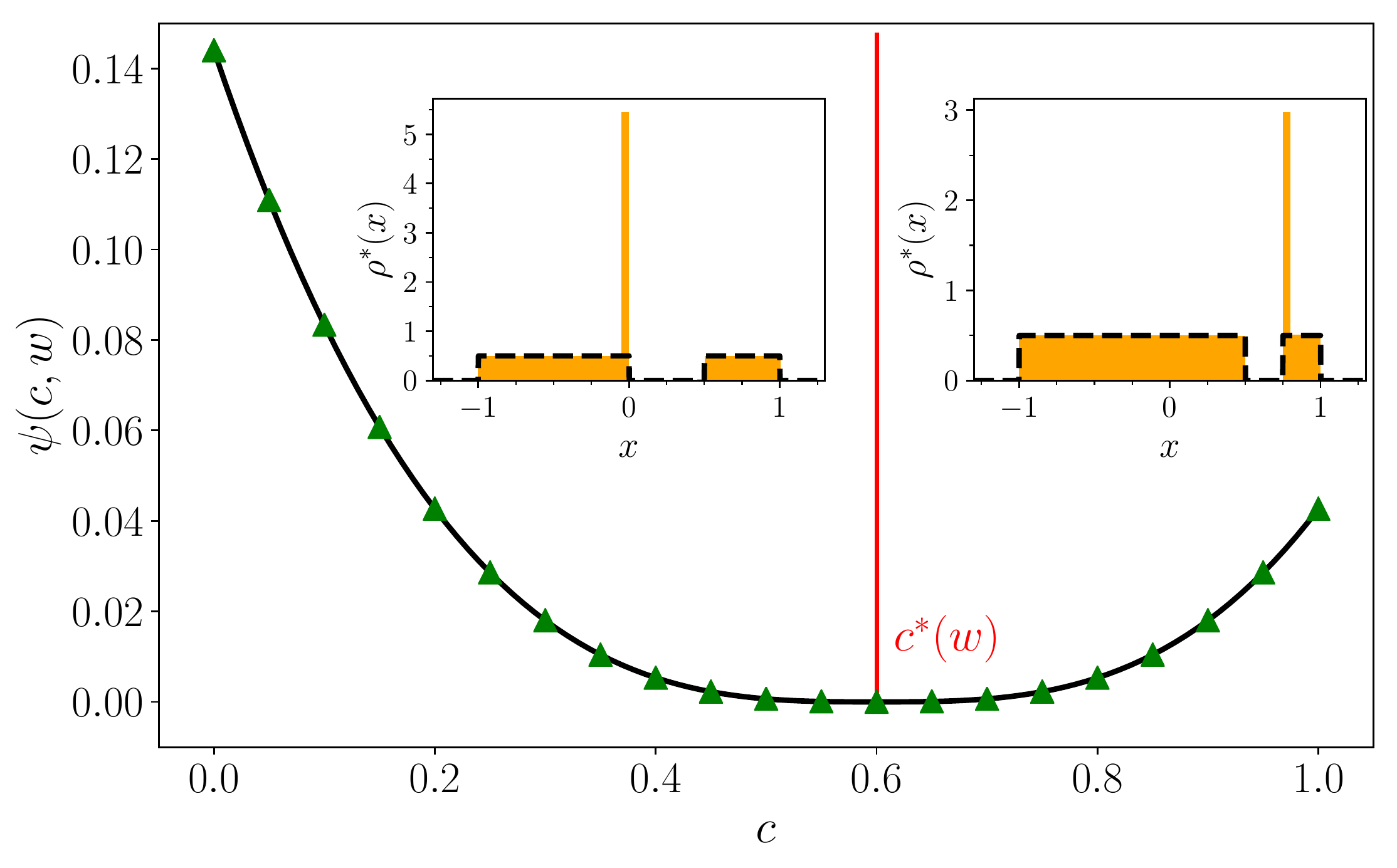}\\
\includegraphics[width=0.494\linewidth, height=4.5cm]{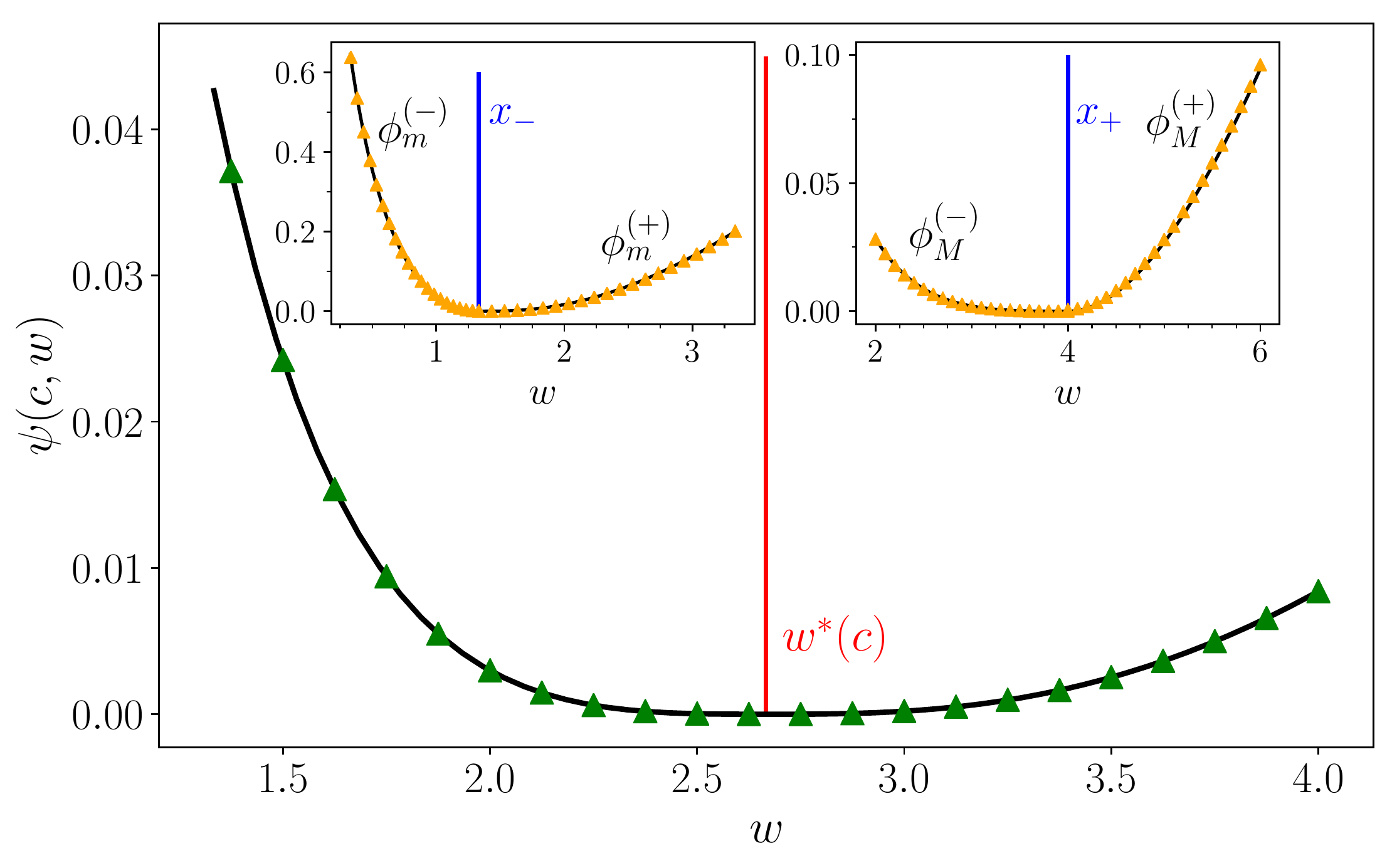}
\includegraphics[width=0.494\linewidth, height=4.5cm]{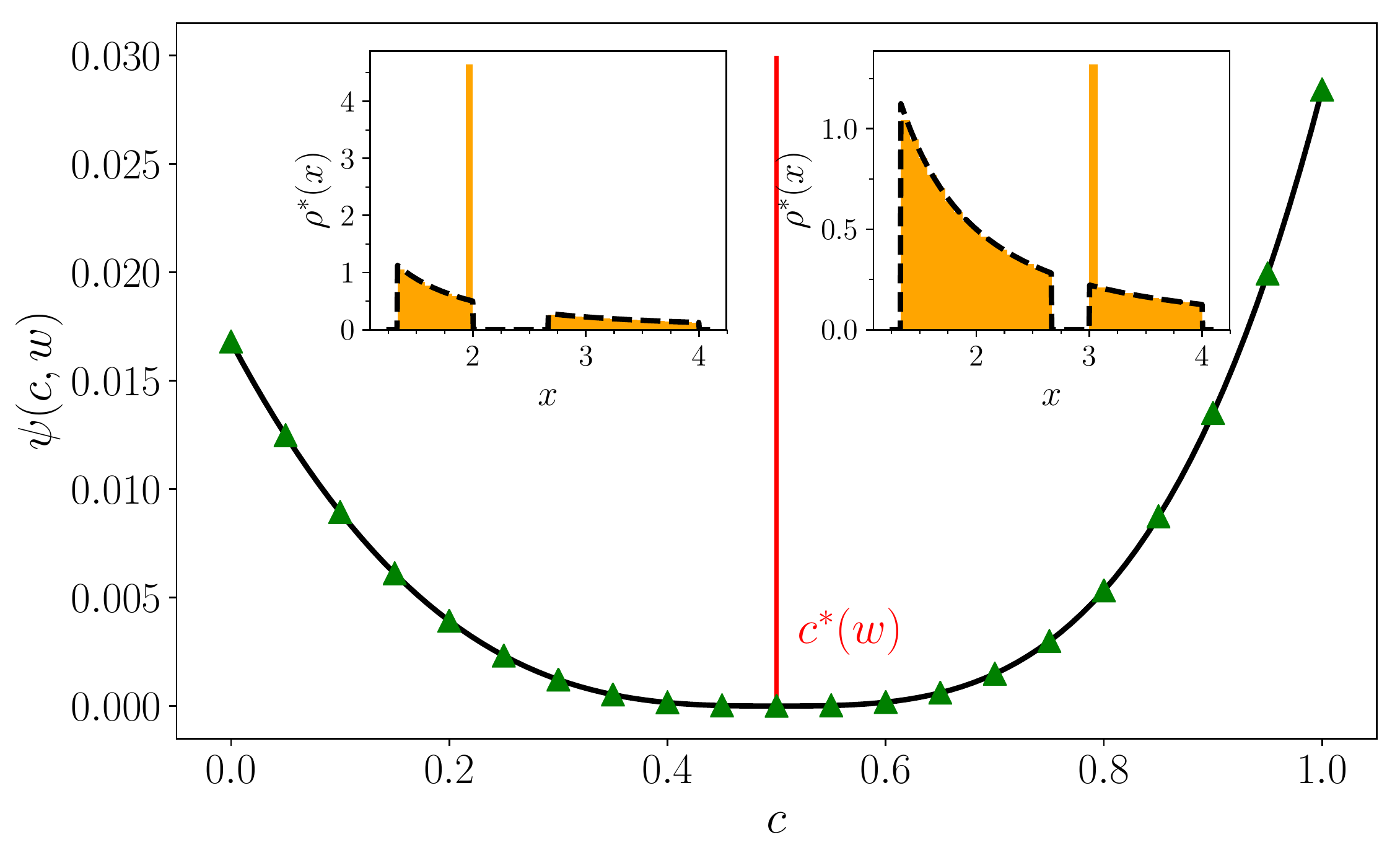}
\caption{
Left panels: Comparison between MC simulations (triangular markers) of $\psi(c,w)$ and the exact expression of Eq.~\eqref{eq:psi general}  (solid curves) as a function of $w$. The upper panel shows the results using the harmonic potential with $\alpha = 1$ and $c=0.5$, while the lower one corresponds to the Wishart-Laguerre potential with $\alpha=0.25$, $a=2$ and $c=0.75$. The right (resp. left) inset shows $\phi_M(w)$  (resp. $\phi_m(w)$), the rate function for $ x_\r{max}$  (resp. $ x_\r{min}$). The MC estimations of, $\phi_M^{(-)}(w)$ and $\phi_m^{{(+)}}(w)$ have been scaled as $N^3$, while the other rate functions were scaled as $N^2$. Right panels: $\psi(c,w)$ as a function of $c$, fixing $w=0.2$ for the harmonic potential (upper panel) and $w=2$ for the Wishart-Laguerre one (lower panel). 
The insets compare MC samples of the constrained density $\rho^*(x)$ (orange histograms) with the exact formula \eqref{eq:density} (blue dashed curve), using $c=0.75$ in all the cases.
}
\label{fig:psi}
\end{figure}

Putting these results together and evaluating Eq.~\eqref{eq:psi actions difference} yields a rather simple expression for the rate function 
\begin{equation}
\begin{split}
\psi(c,w)&= \frac{\abs{c-c^{*}(w)} v(w)}{2} - \int_a^b \dd{x} \frac{v''(x) v(x)}{4\a} \\
&- \frac{\mu^*-\mu^*_0 + \nu^* c}{2} \,,
\label{eq:psi general} 
\end{split}
\end{equation}
whenever the wall is inside the natural support $w\in[x_{-},x_{+}]$ (analogous expressions for $w\not\in[x_{-},x_{+}]$   together with explicit formulas are given in \cite{Note1}). Fig.~\ref{fig:psi} shows a comparison of the analytical value of $\psi(c,w)$ and MC estimations of the rate function for the same two potentials used above.  Importantly, through Eq.~\eqref{eq:psi general} we can recover straightforwardly the results of \cite{Dhar2017,Cunden2017} for the rate functions  $\phi^{(\pm)}_M$ and  $\phi^{(\pm)}_m$  controlling the left and right deviations of the rightmost and leftmost particles (see \cite{Note1} for details).  The left panels' insets of Fig.~\ref{fig:psi} show the comparison of the rate functions of  extremal particles with MC simulations, while the ones in the right panels depict a histogram obtained by MC sampling and the analytical expression for $\rho^*(x)$ according to Eq.~\eqref{eq:density}. In all cases, the agreement is outstanding. Thus Eq.~\eqref{eq:psi general}  provides a general and exact expression for the rate function of one-dimensional CGs and it constitutes the main result of this second part.

As it can be explicitly verified from Eq.~\eqref{eq:psi general} the rate function of the 1d CG has the noticeable feature that its first two partial derivatives vanish at $c^*$ and $w^*$,  where this latter quantity is obtained by inverting the relation defining $c^*$. This is in stark contrast with the analogous result in RMT, where the second partial derivatives are different from zero, meaning that the fluctuations of $x_i$ around $w^*$ (resp.  $\mathcal{C}_w$ around $c^*$) are of Gaussian type  \cite{Orourke2010,Gustavsson2005}. Instead, in the 1d case we found that the \emph{third} derivatives correspond to the first non-vanishing term in the expansion of $\psi(c,w)$ around $c^*$ and $w^*$. In fact, we have that
\begin{equation}\label{eq:asymp w}
\Pr(x_i=w) \asymp \exp( - \frac{N^3 (v''(w^*))^2}{12\a} \abs{w-w^*}^3)\,,
\end{equation}  
which is straightforward to verify that matches exactly with the asymptotic expansion of $F'_{c=iN}(W)$ of Eq.~\eqref{eq:asymptotic Fc}. This last expression implies that the rate function is \emph{not} analytical around its minimum, which is flatter than a quadratic one because of the vanishing second derivatives. We thus end up with the unusual case of having a rate function that is not \emph{strictly} convex nor analytic near its minimum, once again differing from the features of the 2d CG. This is not a minor difference indeed, for it is known \cite{touchette-review-2009,ellis-overview-1995} that a rate function that is not strictly convex can not be extended to the regime of typical fluctuations as in 2d case \cite{Perez2016}. In other words, the 1d CG follows a \emph{weak} large deviations principle, for the rate function cannot be expanded to match smoothly the typical fluctuations regime\cite{touchette-review-2009}. A similar behavior has been found in the 2D Ising system \cite{ioffe-large-deviations-ising,kastner-phase-transition-2002,touchette-review-2009} as well as for a drifted Brownian motion \cite{touchette-minimal}. In \cite{Note1} we provide further evidence that $\psi(c,w)$ does not provide the correct description in the regime of typical fluctuations. As a final remark, our results showed that the rate function for bulk particles in 1d CG exhibits a discontinuity in its third derivate and, while analogous results for extremal particles seem to indicate the presence of a phase transition,  this feature does not  necessarily carry over for bulk particles as fluctuations behave in the same way at each  side of the optimal value $x^*$, as can be observed in Eqs.~\eqref{eq:asymptotic Fc} and \eqref{eq:asymp w}. Thus a phase transition for bulk particles, if any, must lie on another explanation.

\acknowledgements
We thank H. Touchette for helpful comments and for directing us to relevant references about other examples of non analyticity of the rate function. RDHR acknowledges financial support from the London Mathematical Laboratory for performing this research. This work has been funded by the programs UNAM-DGAPA-PAPIIT IA101815 and UNAM-DGAPA-PAPIIT IA103417.

\bibliography{biblio.bib}

\end{document}